\documentclass[]{iopart}
\usepackage{graphicx}
\usepackage{iopams}
\usepackage{color}
\usepackage{bm} 
\usepackage{bbm, dsfont} 
\usepackage{subfigure}
\usepackage{mathrsfs}
\usepackage{verbatim}
\usepackage{cite}

\begin{document}

\title{Selective transport of atomic excitations in a driven chiral-coupled atomic chain}

\author{H. H. Jen}
\address{Institute of Physics, Academia Sinica, Taipei 11529, Taiwan}
\ead{sappyjen@gmail.com}
\def\q{\mathbf{q}}
\renewcommand{\k}{\mathbf{k}}
\renewcommand{\r}{\mathbf{r}}
\newcommand{\parallelsum}{\mathbin{\!/\mkern-5mu/\!}}
\def\p{\mathbf{p}} 
\def\R{\mathbf{R}}
\def\bea{\begin{eqnarray}}
\def\eea{\end{eqnarray}}
\begin{abstract}
We theoretically investigate the flow of the atomic excitations in a driven chiral-coupled atomic chain with nonreciprocal decay channels. This one-dimensional system allows infinite-range dipole-dipole interaction, and enables directional guided modes of scattered light. Under a weakly driven condition, we are able to simulate the transport properties of atomic excitations between the left and right parts of the chain. In the steady states, the transport is highly dependent on the equidistant positions of the ordered array, the excitation field detunings, and the directionality of such chiral-coupled system. We discuss the parameter regimes which are resilient or sensitive to position fluctuations, and provide insights to precise and genuine state preparations. Furthermore, we study the effect of position fluctuations on the transport of excitations. Our results can shed light on deterministic state preparations and pave the way for many-body state manipulations in the driven and dissipative chiral-coupled systems.
\end{abstract}
\maketitle
\section{Introduction}

Manipulating light-matter interactions by precisely positioning quantum emitters has made progress in versatile platforms, including photonic crystal waveguide \cite{Goban2015}, optical microtraps \cite{Barredo2016, Endres2016}, and diamond nanophotonics systems \cite{Sipahigil2016}. This promises to tailor the properties of quantum interface from the bottom up, which is in stark contrast to the atomic ensemble of many randomly distributed emitters \cite{Hammerer2010}. Recently, one-dimensional (1D) atom-nanophotonic waveguide system \cite{Kien2005, Kien2008, Tudela2013, Kumlin2018, Chang2018} presents another potential paradigm to engineer light-matter interactions. This 1D coupled system features strong and infinite-range couplings in the resonant dipole-dipole interactions (RDDI) \cite{Stephen1964, Lehmberg1970}, which is difficult to reach in a free-space atomic system. Only recently, superradiance \cite{Dicke1954, Gross1982} is observed in two atomic clouds above the nanofibers \cite{Solano2017}, demonstrating the infinite-range RDDI in the 1D atom-fiber coupled system. This system is also proposed to realize mesoscopic entangled states \cite{Tudela2013} and allow universal atomic dynamics \cite{Kumlin2018}.  

In addition to the advantage of strong coupling in the 1D atom-fiber or atom-waveguide systems, it can further construct a chiral quantum network \cite{Stannigel2012, Ramos2014, Pichler2015, Lodahl2017}, which enables nonreciprocal decay channels and directional emissions. This chiral coupling breaks the time-reversal symmetry that should be preserved in conventional light-matter interacting systems in free space, and emerges due to the locking of transverse spin angular momentum and light propagation direction \cite{Bliokh2014, Bliokh2015}. The chirality can be engineered via either controlling the atomic internal states \cite{Mitsch2014} or applying external magnetic fields \cite{Luxmoore2013, Sollner2015}. This 1D chiral-coupled system can be used as complementary single-photon devices which are fundamental in quantum internet \cite{Kimble2008}, and can potentially operate CNOT gates \cite{Sollner2015} which are essential to quantum computation. 

In such dynamical system of chiral-coupled atomic chain, the steady-state preparations should highly depend on the positions of the atoms, the driven field detunings, and its directionality of the decay channels. However, the interplay or competition between these parameters is less explored. Here we investigate the selective and controllable transport of atomic excitations to locate the parameter regimes resilient to position fluctuations, which are advantageous to precise and genuine state preparations. The effect of position fluctuations is also discussed, and it shows that the system functions more stable with less than $1\%$ fluctuations. Our results demonstrate potentially deterministic state preparations in the driven and dissipative system and hold promises to manipulate many-body spin dynamics \cite{Hung2016}.  

\begin{figure}[t]
\centering
\includegraphics[width=8.0cm,height=6cm]{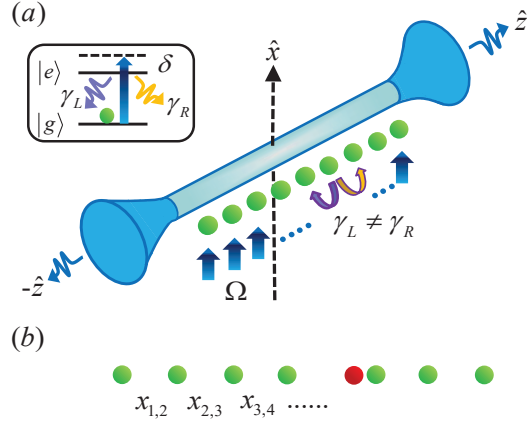}
\caption{Schematic driven chiral-coupled atomic chain. (a) A one-dimensional atom-fiber coupled system shows nonreciprocal decay channels with $\gamma_L\neq\gamma_R$ in $\hat z$. Using weak and uniform external fields $\Omega$ along $\hat x$ with a detuning $\delta$, we are able to drive two-level quantum emitters ($|g(e)\rangle$ for ground and excited states respectively) toward some collective atomic states. This scheme utilizes either the ordered atomic chain with equidistant separations or controllable $x_{i,j}$ between the $i$th and $j$th atoms in (b). A displaced atom in red represents a local disorder, which can further manipulates the atomic steady states via adjusting the collective dipole-dipole interactions in this driven chiral-coupled atomic system.}\label{fig1}
\end{figure}

\section{Effective chiral-coupled interactions}

For atoms in free space, the spontaneous emissions from excited atoms initiate from system-reservoir interactions, which decay as an exponential function with a characteristic time constant. This decay rate represents an intrinsic property of the independent atoms, which can be hugely modified when atoms are close to each other within a transition wavelength. In this regime, strongly-coupled RDDI show up since light can rescatter multiple times before leaving the whole medium. RDDI in a three-dimensional (3D) reservoir are responsible for the collective phenomena of superradiance \cite{Dicke1954, Lehmberg1970, Gross1982, Mazets2007, Jen2012, Jen2015, Sutherland2016_2, Araujo2016, Roof2016, Jennewein2016, Bromley2016, Zhu2016, Shahmoon2017} and subradiance \cite{Scully2015, Guerin2016, Facchinetti2016, Jen2016_SR, Jen2016_SR2, Sutherland2016, Bettles2016, Jen2017_MP, Jenkins2017, Garcia2017, Plankensteiner2017, Jen2018_SR1, Jen2018_SR2} in a dense atomic system. 

RDDI in a 3D reservoir in general are reciprocal couplings which preserve the time-reversal symmetry, and in the long range behave asymptotically as the inverse of mutual atomic separation (see appendix A). By contrast for RDDI in a 1D reservoir, it shows infinite-range couplings in sinusoidal forms \cite{Tudela2013, Kumlin2018}, and can be further structured \cite{Scelle2013, Chen2014, Ramos2014} to show nonreciprocal decay channels in the atom-fiber or atom-waveguide coupled systems. Here we consider a driven chiral-coupled atomic chain as shown in figure \ref{fig1}, where the system dynamics can be described by the effective chiral master equation \cite{Pichler2015}, 
\bea
\frac{d \rho}{dt}=-\frac{i}{\hbar}[H_S+H_L+H_R,\rho]+\mathcal{L}_L[ \rho]+\mathcal{L}_R[ \rho].
\eea
The external light-matter interaction is 
\bea
H_S\equiv \hbar\Omega\sum_{\mu}^N\left(\sigma_\mu+\sigma_\mu^\dag\right)-\hbar\sum_\mu^N\delta_\mu\sigma_\mu^\dag\sigma_\mu,
\eea
which drives the system of $N$ two-level quantum emitters ($|g(e)\rangle$ for the ground and excited state respectively and $\sigma_\mu^\dag\equiv|e\rangle_\mu\langle g|$, $\sigma_\mu=(\hat\sigma_\mu^\dag)^\dag$) with a uniform Rabi frequency $\Omega$ and spatially dependent detunings $\delta_\mu$. The coherent parts are 
\bea
H_L&\equiv&-\frac{i\hbar\gamma_L}{2}\sum_{\mu<\nu}^N\left(e^{ik|x_\mu-x_\nu|}\sigma_\mu^\dag\sigma_\nu-\textrm{H.c.}\right),\\
H_R&\equiv&-\frac{i\hbar\gamma_R}{2}\sum_{\mu>\nu}^N\left(e^{ik|x_\mu-x_\nu|}\sigma_\mu^\dag\sigma_\nu-\textrm{H.c.}\right),
\eea
which denote the collective energy shifts, and the Lindblad forms of 
\bea
\mathcal{L}_L[\hat \rho]\equiv&-\frac{\gamma_L}{2}\sum_{\mu,\nu}^N\left[e^{-ik(x_\mu-x_\nu)}\left(\sigma_\mu^\dag \sigma_\nu \rho +\rho \sigma_\mu^\dag\sigma_\nu-2\sigma_\nu \rho\sigma_\mu^\dag\right)\right],\\
\mathcal{L}_R[\hat \rho]\equiv&-\frac{\gamma_R}{2}\sum_{\mu,\nu}^N\left[e^{ik(x_\mu-x_\nu)}\left(\sigma_\mu^\dag \sigma_\nu \rho +\rho \sigma_\mu^\dag\sigma_\nu-2\sigma_\nu \rho\sigma_\mu^\dag\right)\right],
\eea
characterize the collective decay behaviors. $k=2\pi/\lambda$ is the wave vector for the transition wavelength $\lambda$, and the subscripts $L$ and $R$ respectively label the left- and right-propagating decay channels. For a 1D atomic chain, we can order them as $x_1<x_2<...<x_{N-1}<x_N$ for convenience. The above Lindblad forms do not include the non-guided decay or other non-radiative losses of the atoms, which would compromise the light detection efficiency via fibers or the fidelity of state preparations.

Next we further assume a weakly driven system with $N$ atoms, such that the Hilbert space can be confined in the ground $|g\rangle^{\otimes N}$ and singly excited states $|\psi_\mu\rangle=(\sqrt{N})^{-1}\sum_{\mu=1}^N\sigma_\mu^\dag|g\rangle^{\otimes N}$. This is similar to the coherent dipole model with weak laser excitations \cite{Sutherland2016_2} or Green's function approach in low saturation regime \cite{Tudela2017}, where the ground state population is much larger than the one of the excited state, that is $\langle\sigma_\mu\sigma_\mu^\dag\rangle\approx 1\gg\langle\sigma^\dag_\mu\sigma_\mu\rangle$. In this limit, the state of the system can be expressed as
\bea
|\Psi(t)\rangle=\frac{1}{\sqrt{1+\sum_{\mu=1}^N|A_\mu(t)|^2}}\left(|g\rangle^{\otimes N}+A_\mu(t)|\psi_\mu\rangle\right),
\eea
where the probability amplitude $A_\mu(t)$ can be obtained by
\bea
\dot{A}_\mu(t)=i\Omega+\sum_{\nu=1}^N V_{\mu,\nu}A_\nu(t),
\eea 
and the chiral-coupled interaction $V$ reads
\bea
\fl ~~~~~
V=\left[ \begin{array}{ccccc}
    i\delta_1-\frac{\gamma_L+\gamma_R}{2}  & -\gamma_Le^{-ik|x_{1,2}|} & -\gamma_Le^{-ik|x_{1,3}|} & \dots & -\gamma_Le^{-ik|x_{1,N}|}\\
    -\gamma_Re^{-ik|x_{1,2}|} & i\delta_2-\frac{\gamma_L+\gamma_R}{2} & -\gamma_Le^{-ik|x_{2,3}|} & \dots & -\gamma_Le^{-ik|x_{2,N}|}\\
    -\gamma_Re^{-ik|x_{1,3}|} & -\gamma_Re^{-ik|x_{2,3}|} & i\delta_3-\frac{\gamma_L+\gamma_R}{2} & \dots & -\gamma_Le^{-ik|x_{3,N}|}\\
		\vdots 							& \vdots  & \vdots & \ddots & \vdots  \\
		-\gamma_Re^{-ik|x_{1,N}|} & -\gamma_Re^{-ik|x_{2,N}|} & -\gamma_Re^{-ik|x_{3,N}|} & \dots & i\delta_N-\frac{\gamma_L+\gamma_R}{2}
\end{array}\right],
\eea
where $x_{\mu,\nu}\equiv x_\mu-x_\nu$.

The above chiral-coupled interaction becomes reciprocal, which is $V_{\mu,\nu}=V_{\nu,\mu}$, only when $\gamma_L=\gamma_R$. For reciprocal interaction, $-\gamma_{L(R)}\cos(k|x_{\mu,\nu}|)$ and $\gamma_{L(R)}\sin(k|x_{\mu,\nu}|)$ respectively represent the collective energy shifts and decay rates. In general, $VV^\dag\neq V^\dag V$, so $V$ is not a normal matrix. Therefore, the eigen-decomposition does not work here, and we solve for the system evolutions directly from 
\bea
\frac{d}{dt}\vec A=i\Omega + V \vec A,\label{A}
\eea
where $\vec A\equiv (A_1(t),A_2(t),...,A_N(t))$ with the initial conditions of $\vec A(t=0)=0$. Below we use equation (\ref{A}) to investigate the transport properties of atomic excitations and state manipulations in a driven chiral-coupled atomic chain.
 
\section{Selective transport of atomic excitations}

Here we quantify the transport of atomic excitations by the difference of excited state populations between the left and right sections of the atomic chain for even and odd $N$ respectively,  
\bea
T_p = \frac{\sum_{\mu=1}^{N/2,(N-1)/2}P_\mu(\infty) - \sum_{\mu=N/2+1,(N+3)/2}^N P_\mu(\infty)}{\sum_{\mu=1}^N P_\mu(\infty)},
\eea
where the excited state population is $P_\mu\equiv|A_\mu|^2$, and we have excluded the central atom (the $[(N+1)/2]$th one) for odd $N$. Positive or negative $T_p$ means that the atomic excitations accumulate toward the left or right parts of the chain, from which we can analyze how the distributions of the excitations are manipulated and controlled by system parameters. $T_p$ can be used as indicators of how efficient the light is transferred to either directions of the chain, and can further apply to quantum links between multiple atomic chains. 

\subsection{Two atoms}

Before we study a longer atomic chain, it is helpful to study the case of $N=2$, where we can get some insights of how atomic excitations are distributed. The coupled equations from equation (\ref{A}) are
\bea
\dot{A}_1(t)=&i\Omega+\left(i\delta_1-\frac{\gamma}{2}\right)A_1-\gamma_L e^{-i\xi}A_2,\\
\dot{A}_2(t)=&i\Omega - \gamma_Re^{-i\xi}A_1 + \left(i\delta_2-\frac{\gamma}{2}\right)A_2,
\eea
where a dimensionless $\xi\equiv k|x_1-x_2|$ sets the length scale for mutual separations. In the steady state, we obtain 
\bea
A_1(\infty) =& \frac{-i\Omega\left(i\delta_2-\gamma/2+\gamma_Le^{-i\xi}\right)}{(i\delta_1-\gamma/2)(i\delta_2-\gamma/2)-(\gamma-\gamma_L)\gamma_Le^{-i2\xi}},\\
A_2(\infty) =& \frac{-i\Omega\left[i\delta_1-\gamma/2+(\gamma-\gamma_L)e^{-i\xi}\right]}{(i\delta_1-\gamma/2)(i\delta_2-\gamma/2)-(\gamma-\gamma_L)\gamma_Le^{-i2\xi}},
\eea
where we have replaced $\gamma_R=\gamma - \gamma_L$ with $\gamma$ indicating the overall decay rate of the system. The steady-state populations are thus determined by individual detunings, mutual distances, and the directionality factor $D\equiv (\gamma_R-\gamma_L)/\gamma$ \cite{Mitsch2014}. $D=\pm 1$ means the cascaded scheme \cite{Stannigel2012, Gardiner1993, Carmichael1993} where only the guided emission to the right or left is allowed, while $D=0$ represents the reciprocal case as in conventional light-matter interacting systems without the presence of nanofiber or waveguide.

\begin{figure}[t]
\centering
\includegraphics[width=12.0cm,height=6cm]{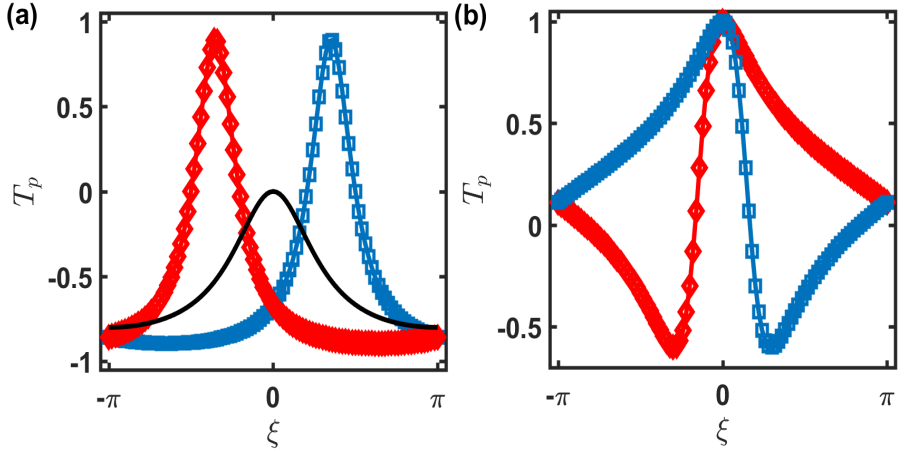}
\caption{Distribution of atomic excitations for $N=2$. (a) The cascaded case ($D=1$) with $\delta_2/\gamma=0$ and $\delta_1/\gamma=-1$ (red-$\diamond$), $0$ (solid-black), $1$ (blue-$\square$), respectively. (b) The case for reciprocal couplings with $\gamma_L=\gamma_R$ ($D=0$) is shown with $\delta_1/\gamma=0$ and $\delta_2/\gamma=$ $-0.5$ (red-$\diamond$), $0.5$ (blue-$\square$), respectively.}\label{fig2}
\end{figure}

For $N=2$, we obtain
\bea
T_p(N=2) = \frac{|i\delta_2-1/2+\gamma_Le^{-i\xi}|^2-|i\delta_1-1/2+(1-\gamma_L)e^{-i\xi}|^2}{|i\delta_2-1/2+\gamma_Le^{-i\xi}|^2+|i\delta_1-1/2+(1-\gamma_L)e^{-i\xi}|^2}.
\eea
In figure \ref{fig2}, we plot the distribution of atomic excitations for $N=2$. For the cascaded scheme of $D=1$ in figure \ref{fig2}(a), the mirror symmetry shows in the cases of $\pm\delta_1$ at $\xi=0$. Since $D=1$ means unidirectional decay channel to the right, a preference of the excitation transfer to the right (negative $T_p$) should cover most of the parameter regimes of $\xi$. In this cascaded scheme, a relatively sharp profile shows up around $\xi\sim\mp\pi/3$ for a positive $T_p$, while a flattened one for a negative $T_p$ at $\xi\sim\pm\pi/2$, respectively for $\delta_1=\mp 1\gamma$. The positive transport in this cascaded scheme manifests the dominance of 1D RDDI over the directionality of the chiral-coupled system. Meanwhile, the flattened regions of $T_p$ indicates a resilient response of position fluctuations in $\xi$, which is preferential for precise and genuine state preparations.

On the other hand for reciprocal couplings of $D=0$ in figure \ref{fig2}(a), similar asymmetric profile emerges and shows mirror reflection at $\xi=0$ with $\pm\delta_2$. Specifically at $\xi=0$, $T_p$ becomes one as long as the left atom is resonantly driven, which means a complete suppression of the atomic excitations on the right. This dispersion-like distribution of atomic excitations enables both positive and negative transport around $\xi=0$, making flexible steady-state preparations simply by either manipulating excitation detunings or $\xi$. Under the resonance condition $\delta_{1(2)}=0$, $T_p$ is not defined since $A_{1(2)}(\infty)$ becomes infinite. This is due to the breakdown of the low saturation approximation in the model, which will be addressed in the end of the next subsection. Next we study a longer atomic chain and investigate the multi-atom effect on the transport property.

\subsection{Atomic chain}

For a longer atomic chain, the left- and right-propagating emissions can go through multiple scatterings of transmissions and reflections before leaving the whole array. As more atoms are included, the many-body coherences will play important roles in determining the steady-state properties. Further with varying equidistant positions $\xi$, the chiral-coupled interactions can also be modified significantly. Therefore, here we focus on the interplay between the number of atoms and their positions in the transport of atomic excitations. 

\begin{figure}[t]
\centering
\includegraphics[width=12.0cm,height=6cm]{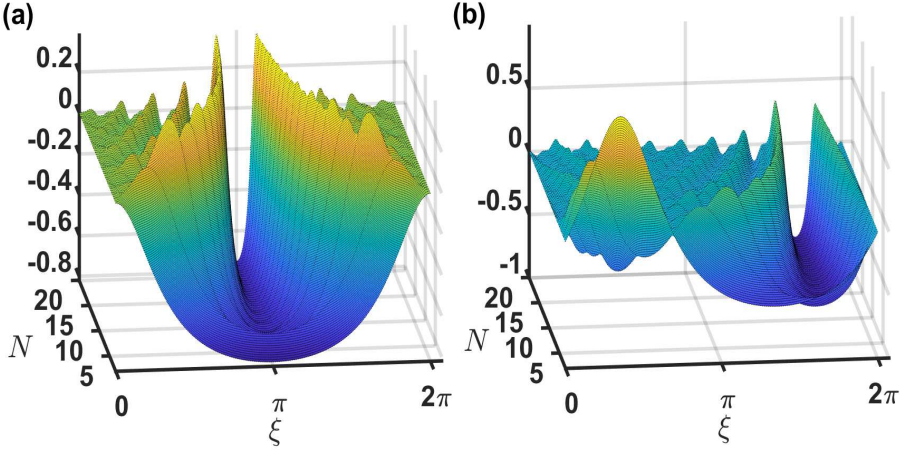}
\caption{The transport $T_p$ of atomic excitations as a dependence of $N$ and $\xi$ for $D=1$. With uniform detunings (a) $\delta_\mu/\gamma=0$ and (b) $\delta_\mu/\gamma=1$, $T_p$ shows positive or negative flow of atomic excitations up to $N=20$. We note that $\xi\in[0,2\pi]$ is periodic by $2\pi$, and thus is equivalent to $\xi\in[-\pi,\pi]$.}\label{fig3}
\end{figure}

First we consider the cascaded case ($D=1$) with uniform detunings in figure \ref{fig3}. On resonance, the transport profiles should be symmetric around $\xi=0$ or $\pi$ as shown in figure \ref{fig3}(a), similar to figure \ref{fig2}(b) for $N=2$. As $N$ increases, the width of the minimum $T_p(\xi=\pi)$ narrows, and positive $T_p$ becomes sharper near $\xi\sim\pi$. More ripples around $T_p\approx 0$ show up as $N$ increases, which indicates of multiple interferences from these quantum emitters. This also appears in figure \ref{fig3}(b) with finite excitation detunings, where the minimum of $T_p$ shifts toward $\xi=2\pi$. This suggests to allow adjustable transport of excitations by controlling external fields and locating the optimal $\xi$. The narrowing distribution of $T_p$ for larger $N$, however, restricts a genuine preparation of the states if the atomic chain undergoes significant position fluctuations. For an example of $N=10$ in figure \ref{fig3}(a), the full width of the half minimum $T_p(\xi=\pi)$ is $\Delta\xi\sim 1$, which provides an approximate tolerance of position fluctuations around $\pm 0.5/\pi$, i.e. $\sim\pm 15\%$ displacement around $\xi=\pi$. Thus for an atomic chain of $N>10$ subject to more significant fluctuations $\gtrsim 15\%$, it is more demanding in stabilizing the system to transfer the excitations with high fidelity.

\begin{figure}[b]
\centering
\includegraphics[width=12.0cm,height=6cm]{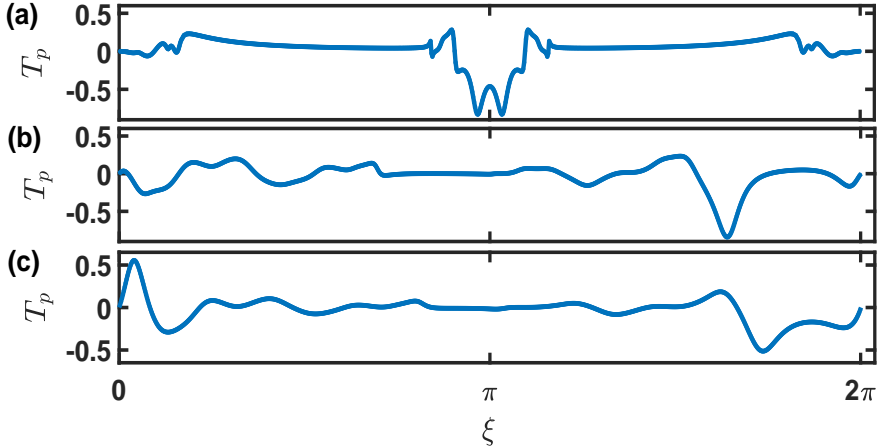}
\caption{The transport $T_p$ of atomic excitations for $N=10$ and $D=0.5$. Symmetric and asymmetric profiles of $T_p$ are plotted for various uniform detunings in (a) $\delta_\mu/\gamma=0$, (b) $\delta_\mu/\gamma=1$, and (c) $\delta_\mu/\gamma=2$, respectively.}\label{fig4}
\end{figure}

On the other hand for the case of nonreciprocal decay channels, we consider $D=0.5$ as an example. We show $T_p$ in figure \ref{fig4} for $N=10$ with increasing detunings. Similar to the cascaded scheme of figure \ref{fig3}(a), in figure \ref{fig4}(a), the transport of atomic excitations is symmetric at $\xi=0$ or $\pi$ under the conditions of resonantly driven fields. The small bump at $\xi=\pi$ emerges as long as $D<1$, which can be further enhanced as $D$ decreases. Other than the parameter regime of $\xi\sim\pi$, $T_p$ shows relatively flattened distributions with almost equal excitation populations between the left and right parts of the chain. By contrast when detunings are increasing in figures \ref{fig4}(b) and (c), the minimum of $T_p$ shifts toward $\xi=2\pi$. Significant positive $T_p$ at $\xi\gtrsim 0$ emerges as well in figure \ref{fig4}(c), which is over $0.5$. Furthermore in figure \ref{fig4}(c), the width of this negative peak of $T_p$ widens as $\delta_\mu$ increases. This shows the possibility, in the widened regions of $T_p$, to manipulate a genuine transport of atomic excitations in the nonreciprocal scheme. In addition, the red-detuned excitation fields hold a symmetric relation in the transport property, such that $T_p(\delta_\mu=-\delta, -\xi)=T_p(\delta_\mu=\delta, \xi)$.

We note that as $D\rightarrow 0$ and $N\gg 1$, the time to reach steady states prolongs and even longer at $\xi\sim\pi$. At the special parameter regime of $D=0$, $\delta_\mu=0$, and $\xi=\pi$, the excited atoms under reciprocal couplings become decoherence-free, and thus are pumped by external fields $\Omega$ indefinitely. This eventually violates the assumptions of small excited state populations in our models in section 2. The validity of weakly-coupled regime can be retrieved with a finite detuning $\delta_\mu=\delta$ and satisfying $\Omega/\delta\ll 1$. Under this condition, the system evolves with a generalized Rabi frequency $\sim \delta$, and exchanges between the ground and the singly-excited states, where the steady-state is never reached.


\section{Effect of position fluctuations of the atomic chain}

\begin{figure}[t]
\centering
\includegraphics[width=12.0cm,height=6cm]{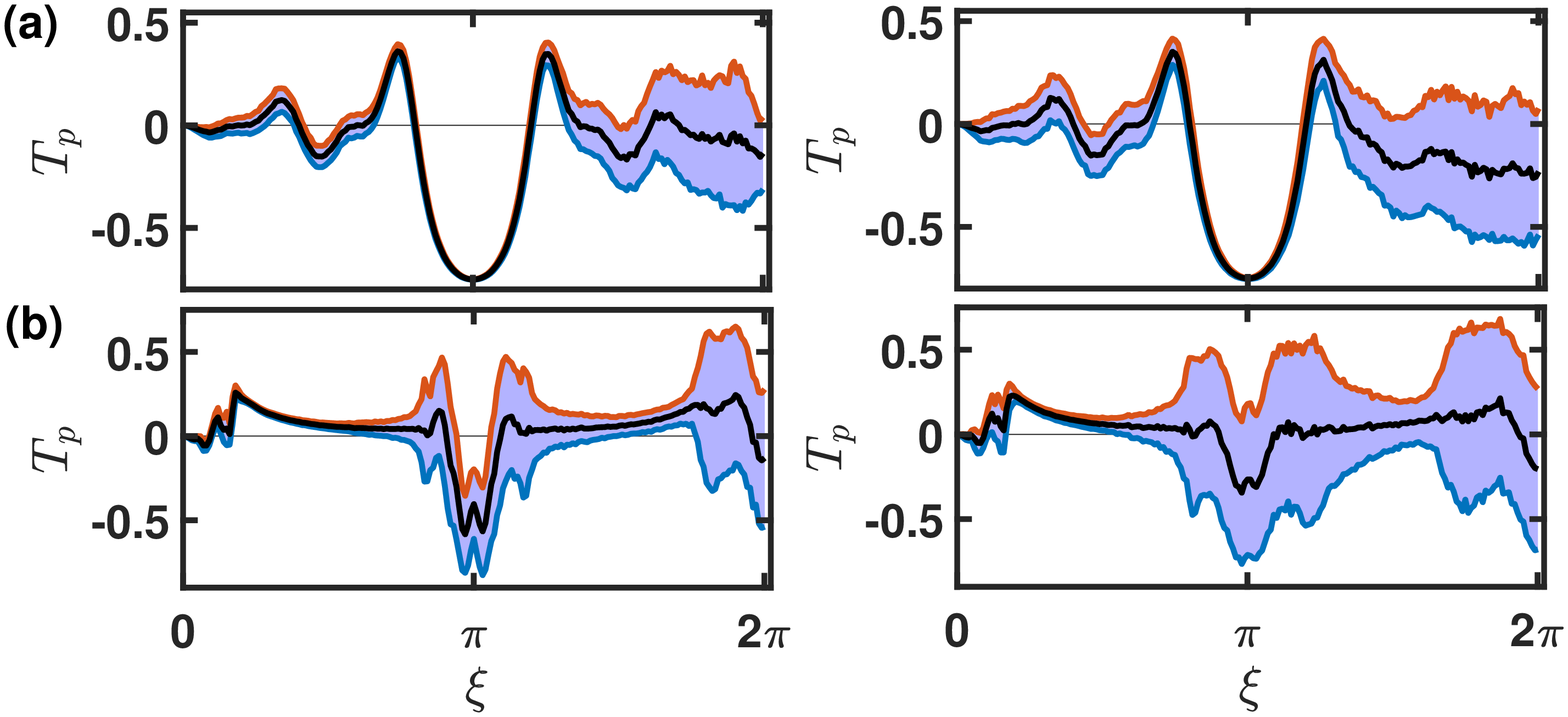}
\caption{The transport $T_p$ of atomic excitations for $N=10$ and $\delta_\mu=0$. (a) The cascaded case of $D=0$ and (b) the case with nonreciprocal decay channels of $D=0.5$, with $0.5\%$ (left) and $1\%$ (right) position fluctuations respectively. Shaded areas are filled between the upper and lower curves with $1\sigma$ standard deviation, and a solid line in black represents the mean value with ensemble averages.}\label{fig5}
\end{figure}

Finally we study the effect of position fluctuations on the transfer of state excitations in the driven and chiral-coupled atomic chain. On the experimental side, a precise positioning of the atoms is not easily fulfilled. Whether the spatial variations of the atoms make a notable effect or not depends on the ratio of deviation and the transition wavelength. Superconducting qubits, for example, is more resilient to this spatial fluctuation due to microwave transmission line, in contrast to the optical transition of neutral atoms.

In figure \ref{fig5}, we plot the $T_p$ by introducing the position fluctuations on each atoms of the chain. As the fluctuation increases, $T_p$ smooths out especially close to $\xi\sim 2\pi$, compared to figures \ref{fig3}(a) and \ref{fig4}(a). $T_p$ is more sensitive to the spatial variation at large $\xi$ since the change of $e^{ik|x_\mu-x_\nu|}$ in the RDDI is more radical under the same amount of position fluctuations. For the cascaded case of $D=1$ in figure \ref{fig5}(a) near $\xi\sim\pi$, the transport of the atomic excitations is less affected by the fluctuations, in contrast to the non-cascaded case of $D=0.5$ in figure \ref{fig5}(b). This is due to the fact that the non-cascaded scheme permits both left- and right-propagating decay channels, and therefore the driven atomic chain experiences the position fluctuations more significantly from both directions. This shows that the cascaded chiral-coupled atomic chain can better withstand the influences of atomic spatial fluctuations. 

\section{Discussion and conclusion}

The efficiency and fidelity of state preparations can be reduced due to non-guided modes of light in the chiral-coupled systems. This limits the operation time in state manipulations, and thus the non-guided decay rate sets the overall timescale for genuinely controlled dynamical systems. The inefficiency can be overcame, for example in an atom-fiber system, by aligning the atoms close to the nanofiber with an optimal fiber radius \cite{Kien2005_2, Chang2018} to raise the interaction probability. Some recent progress has shown the potential of flexible control over the chiral-coupled systems, including the superconducting qubits without external magnetic fields \cite{Hamann2018} and a bilayer atomic array in free space \cite{Grankin2018}. This shows rich opportunities in structuring the 1D reservoirs to manipulate the directionality of light coupling in the system. Moreover, subradiance dynamics can also emerge in such chiral-coupled atomic chain \cite{Jen2018_chiral}, which can be potentially used for quantum storage of light. With the scalability of atom-fiber or atom-waveguide platforms, the chiral-coupled systems can further stimulate applications in quantum information processing.

In conclusion, we have investigated the distribution of the atomic excitations in a weakly driven chiral-coupled atomic chain. In this 1D system, we quantify the distributions by the transport of the excitations between the left and right parts of the chain. With controllable parameters of the external field detunings, ordered array positions, and directionality of the decay channels, we are able to make a deterministic transfer of atomic excitations to both sides. With the advantage of tunable nonreciprocal couplings, chiral-coupled system allows a selective transport of excitations in the optimal parameter regimes resilient to position fluctuations. Our results provide a fundamental study on the steady-state preparations in the driven and dissipative chiral-coupled systems, and are potentially applicable in many-body state manipulations. 

\appendix
\section{Resonant dipole-dipole interactions in one-dimensional reservoir}

\subsection{General formalism in three-dimensional reservoir}

Here we review the results of resonant dipole-dipole interactions (RDDI) in one-dimensional (1D) reservoir. This can be directly obtained and extended from the RDDI in three-dimensional (3D) reservoir \cite{Stephen1964, Lehmberg1970}. The RDDI result from a system of many atoms interacting with quantized bosonic light modes. Due to the light rescattering events in the dissipation process of the system, the RDDI that feature long-range atom-atom interactions emerge as if the whole medium are effectively resonantly induced dipoles in pairs. The RDDI characterize the coherent frequency shifts and collective decay rates of the atomic system, which can be expressed respectively as imaginary and real parts of the coupling constant $J_{\mu,\nu}$, such that the evolutions of any atomic observables $Q$ can be governed by Lindblad forms,
\bea
\dot{Q}(t)&=i\sum_{\mu\neq\nu}{\rm Im}(J_{\mu,\nu})[\sigma_\mu^\dag\sigma_\nu,Q]+\mathcal{L}(Q),\\
\mathcal{L}(Q)&=\sum_{\mu,\nu}{\rm Re}(J_{\mu,\nu})\left[2\sigma_\mu^\dag Q\sigma_\nu-(\sigma_\mu^\dag\sigma_\nu Q+Q\sigma_\mu^\dag\sigma_\nu)\right],
\eea 
where $\sigma_\mu^\dag\equiv|e\rangle_\mu\langle g|$ and $\sigma_\mu=(\hat\sigma_\mu^\dag)^\dag$ are raising and annihilating operators respectively for ground $|g\rangle$ and excited states $|e\rangle$. The above form can be obtained with the Born-Markov and secular approximations, which can be sustained respectively when the response time of the reservoir is faster than the system and the dynamical time scale of the system is longer than the time light travels throughout the whole medium. These conditions can be satisfied when the macroscopic length scale of the medium is way below several meters for rubidium atoms (intrinsic decay time $\sim 26$ ns). 

The explicit forms of $J_{\mu,\nu}$ are defined as
\bea
J_{\mu,\nu}&=\sum_q |g_q|^2\int_0^\infty dt' e^{i\k_q\cdot(\r_\mu-\r_\nu)}[e^{i(\omega_e-\omega_q)t'}+e^{-i(\omega_e+\omega_q)t'}], \nonumber\\
&=\sum_q |g_q|^2\int_0^\infty dt' e^{i\k_q\cdot(\r_\mu-\r_\nu)}[\pi\delta(\omega_q-\omega_e)+\pi\delta(\omega_q+\omega_e)\nonumber\\
&+i\mathcal{P}(\omega_e-\omega_q)^{-1}-i\mathcal{P}(\omega_q+\omega_e)^{-1}],\label{J}
\eea
where the coupling constant is $g_q\equiv d/\hbar\sqrt{\hbar\omega_q/(2\epsilon_0V)}(\vec\epsilon_q\cdot\hat d)$ with dipole moment $d$ and its unit direction $\hat d$, field polarizations $\vec\epsilon_q$, and a quantization volume $V$. $\mathcal{P}$ is the principal value of the integral. From equation (\ref{J}), the coupling constant depends on respective atomic positions $\r_{\mu(\nu)}$, and thus is a long-range interaction.

For a 3D reservoir, we allow continuous modes of reservoir, that is $\sum_q\rightarrow\sum_{\vec\epsilon_q}\int_{-\infty}^\infty\frac{V}{(2\pi)^3}d^3q$ with two possible field polarizations $\vec\epsilon_q$. In spherical coordinates, we show $J_{\mu,\nu}$ explicitly \cite{Lehmberg1970},
\bea 
\fl
{\rm Re}(2J_{\mu,\nu})=&\frac{3\Gamma}{2}\bigg\{\left[1-(\hat\p\cdot\hat{r}_{\mu\nu})^2\right]\frac{\sin\xi}{\xi}
+\left[1-3(\hat\p\cdot\hat{r}_{\mu\nu})^2\right]\left(\frac{\cos\xi}{\xi^2}-\frac{\sin\xi}{\xi^3}\right)\bigg\},\label{F}\\
\fl
{\rm Im}(J_{\mu,\nu})=&\frac{3\Gamma}{4}\bigg\{-\Big[1-(\hat\p\cdot\hat{r}_{\mu\nu})^2\Big]\frac{\cos\xi}{\xi}
+\Big[1-3(\hat\p\cdot\hat{r}_{\mu\nu})^2\Big]
\left(\frac{\sin\xi}{\xi^2}+\frac{\cos\xi}{\xi^3}\right)\bigg\}\label{G}, 
\eea
where $\hat\p$ aligns with the excitation field polarization, the natural decay constant $\Gamma=d^2\omega_e^3/(3\pi\hbar\epsilon_0c^3)$, and dimensionless $\xi\equiv k_L|\r_\mu-\r_\nu|$ with $k_L=\omega_e/c$. 

\subsection{Results of RDDI in one-dimensional reservoir}

Now it is straightforward to derive the RDDI in 1D reservoir from equation (\ref{J}). Since the reservoir only allows one dimension, we should replace $V$ in $g_q$ by $L$ as a length scale of the quantization volume. We then obtain 
\bea
J_{\mu,\nu}=&\int_{-\infty}^\infty\frac{dq}{2\pi}\bar g_q^2 L e^{i\k_q\cdot(\r_\mu-\r_\nu)}[\pi\delta(\omega_q-\omega_e)+\pi\delta(\omega_q+\omega_e)\nonumber\\&+i\mathcal{P}(\omega_e-\omega_q)^{-1}-i\mathcal{P}(\omega_q+\omega_e)^{-1}],
\eea
where $\bar g_q^2$ $\equiv$ $(d/\hbar)^2[\hbar\omega_q/(2\epsilon_0 L)]$ and the term $(\vec\epsilon_q\cdot\hat d)$ becomes one due to the spin-momentum locking in 1D reservoir. 

Next we let $x_{\mu,\nu}=x_\mu-x_\nu$ and drop $q$ in $\omega_q$ for brevity, and we obtain
\bea
J_{\mu,\nu}=&\int_0^\infty\frac{d\omega}{\pi}|\partial_\omega q(\omega)|\bar g_q^2 L\cos(\k_qx_{\mu,\nu})[\pi\delta(\omega-\omega_e)+\pi\delta(\omega+\omega_e)\nonumber\\&+i\mathcal{P}(\omega_e-\omega)^{-1}-i\mathcal{P}(\omega+\omega_e)^{-1}].
\eea
Let $\Gamma_{1D}\equiv 2|\partial_\omega q(\omega)|_{\omega=\omega_e}\bar g_{k_L}^2L$, where we keep the dispersion relation of the 1D coupling constant with $\partial_\omega q(\omega)$ being the group velocity of light in the medium. Finally we obtain
\bea
\fl ~~~~~~~~~~
J_{\mu,\nu}&=\frac{\Gamma_{1D}}{2}\cos(k_L x_{\mu,\nu})-\frac{i\mathcal{P}}{\pi}\int_{-\infty}^\infty d\omega
\frac{{\rm Re}[|\partial_\omega q(\omega)|\bar g_q^2Le^{i\k_q(x_\mu-x_\nu)}]}{\omega-\omega_e},\\
\fl
&=\frac{\Gamma_{1D}}{2}\left[\cos(k_L x_{\mu,\nu})+i\sin(k_L |x_{\mu,\nu}|)\right].\label{chiral1D}
\eea
where the respective real and imaginary parts should demonstrate the Kramers-Kronig relation \cite{Zangwill2013}. 

\ack
This work is supported by the Ministry of Science and Technology (MOST), Taiwan, under the Grant No. MOST-106-2112-M-001-005-MY3 and 107-2811-M-001-1524. We thank Y.-C. Chen, G.-D. Lin, and M.-S. Chang for insightful discussions, and are also grateful for NCTS ECP1 (Experimental Collaboration Program).

\end{document}